\documentstyle[12pt,equation]{article}

\begin{document}
\def\to{\rightarrow}
\def\ub{\underbar}
\def\vb{\mbox{$\bar{v}$}}
\def\wt{\widetilde}
\def\ta{\mbox{$\widetilde{A}$}}
\def\ffbar{\mbox{$f \bar{f}$}}
\def\lmax{\mbox{$\lambda_{\rm max}$}}
\def\tanb{\mbox{$\tan \! \beta$}}
\def\sinb{\mbox{$\sin \! \beta$}}
\def\cosb{\mbox{$\cos \! \beta$}}
\def\ctw{\mbox{$\cos \! \theta_W$}}
\def\stw{\mbox{$\sin \! \theta_W$}}
\def\hth{\mbox{$H_3^0$}}
\def\nt{\mbox{$N_3$}}
\def\to{\rightarrow}
\def\hbt{\mbox{$\bar{H^0_3}$}}
\def\ml{\mbox{$m_{L_1^0}$}}
\def\lone{\mbox{$L_1^0$}}
\def\mlone{\mbox{$m_{L_1^0}$}}
\def\ltwo{\mbox{$L_2^0$}}
\def\sigav{\mbox{$\langle \sigma_{\rm eff} v \rangle$}}
\def\omeh{\mbox{$\Omega h^2$}}
\def\mlop{\mbox{$m_{L_1^\pm}$}}
\def\mltp{\mbox{$m_{L_2^\pm}$}}
\def\la{\mbox{$\lambda$}}
\def\rs{\mbox{$\sqrt{s}$}}
\def\qqbar{\mbox{$q \bar{q}$}}
\def\ben{\begin{subequations}}
\def\be{\begin{equation}}
\def\een{\end{subequations}}
\def\ee{\end{equation}}
\def\beq{\begin{eqalignno}}
\def\eeq{\end{eqalignno}}
\def\bea{\begin{eqnarray}}
\def\eea{\end{eqnarray}}
\def\epem{\mbox{$e^+ e^- $}}
\def\aem{\mbox{$\alpha_{\rm em}$}}
\renewcommand{\thefootnote}{\fnsymbol{footnote}}
\setcounter{page}{0}
\setcounter{footnote}{0}
\begin{flushright}
MADPH--95--908\\
September 1995\\
\end{flushright}
\begin{center}
\vspace*{1cm}
{\large \bf Testing Superstring--Inspired $E(6)$ Models at Linear \epem\
Colliders}\footnote{Talk given at the {\it International Workshop on Physics
and Experiments at Linear \epem\ Colliders}, Iwate, Japan, September 1995}\\
\vspace{6mm}
Manuel Drees\footnote{Heisenberg Fellow}\\
\vspace{5mm}
Department of Physics, University of Wisconsin, Madison, WI 53706, USA \\
\vspace{9mm}
\end{center}

\begin{abstract}
It is shown that in a large class of $E(6)$ models, either an \epem\ collider
operating at $\rs \geq 1.5$ TeV must find a signal for the production of
exotic leptons, or a collider operating at $\rs \geq 300$ GeV must find at
least one light neutral Higgs boson with a large invisible branching ratio.
The region of parameter space where neither of these signals is visible can be
excluded because here the lightest neutral exotic lepton, which is absolutely
stable, would overclose the Universe.
\end{abstract}
\newpage
\pagestyle{plain}
\setcounter{page}{1}

Superstring--inspired $E(6)$ models \cite{1} postulate the existence of a
large number of new (s)particles, including new gauge bosons, exotic quarks
and leptons, lepto-- or di--quarks, and an extended Higgs sector.
Unfortunately most of these (s)particles might be very heavy; the structure of
the theory allows to increase their masses more or less arbitrarily, either
``by hand" (as in case of sfermions and gauginos), or by coupling them to the
vev $x$ of an SM singlet Higgs field. Searches for these (s)particles can
therefore constrain (or find evidence for), but never strictly rule out this
class of models.

There are two notable exceptions to this rule, however. One is the lightest
scalar neutral Higgs boson which, as in all weakly coupled supersymmetric
theories \cite{2}, must have \cite{3} a mass below 150 GeV or so. Large lower
bounds \cite{1,4} on the masses of new gauge bosons imply that this light
Higgs boson must be essentially a mixture of the $SU(2)$ doublet Higgs fields
of the model, as in the MSSM; an \epem\ collider operating at center--of--mass
energy $\rs \geq 300$ GeV must therefore find at least one of these Higgs
bosons \cite{5}. However, precisely because all (weakly coupled) SUSY theories
predict \cite{2} the existence of at least one observable \cite{6}, rather
light Higgs boson, this does not lead to a very specific test of the $E(6)$
models under consideration.

Such an $E(6)$ specific test can be devised \cite{7} using the observation
\cite{8} that these models also predict the existence of at least two exotic
neutral ``leptons" with masses not much above that of the $Z$ boson. Each {\bf
27} of $E(6)$ contains five neutral fermions: The usual neutrino $\nu_L$; the
exotic $SU(2)$ doublets $\wt{H^0}, \wt{\bar{H^0}}$; and the Standard Model
singlets $\nu_R$ and $\wt{N}$. $\nu_R$ resides in the {\bf 16} of $SO(10)$; it
might be essentially massless, or it may obtain a mass $\geq 1$ TeV due to
nonrenormalizable interactions of the form $\frac {1}{M_{Pl}} \langle \tilde
\nu^*_R \rangle \langle \tilde \nu^*_R \rangle \nu_R \nu_R$ if $\langle \tilde
\nu_R \rangle \geq 10^{10}$ GeV, as in certain see--saw models of neutrino
masses \cite{9}. In either case the $\nu_R$ fields have little impact on
collider phenomenology.

This is quite different for the remaining three neutral fermions per
generation. It is always possible to choose a basis where only one of the
scalar $N, \ H^0$ and $\bar{H^0}$ fields has a nonzero vev; following the
notation of ref.\cite{10}, these are denoted by $N_3, \ H^0_3$ and
$\bar{H^0_3}$, with $\langle N_3 \rangle \equiv x, \ \langle H^0_3
\rangle\equiv v$, and $\langle \bar{H^0_3} \rangle \equiv \bar v$. The
fermionic superpartners of these Higgs fields then mix with the gauginos of
the model to form the six (or more) neutralino states. Here we are interested
in the first two generations of fermionic $H, \ \bar{H}$ and $N$ fields. They
obtain their masses from the superpotential
\be \label{e1}
W_{\rm lep} = \sum_{i,j,k=1}^3 \la_{ijk} H_i \bar{H}_j N_k.
\ee
This gives rise to the mass matrix [in the basis $(\wt{H_1^0},
\wt{\bar{H^0_1}}, \wt{N_1}, \wt{H_2^0}, \wt{\bar{H^0_2}}, \wt{N_2})$]:
\be \label{e2}
{\cal M}_{L^0} = \mbox{$
\left( \begin{array}{cccccc}
0 & \mlop & \la_{131} \vb & 0 & 0 & \la_{132} \vb \\
\mlop & 0 & \la_{311} v & 0 & 0 & \la_{312} v \\
\la_{131} \vb & \la_{311} v & 0 & \la_{231} \vb & \la_{321} v & 0 \\
0 & 0 & \la_{231} \vb & 0 & \mltp & \la_{232} \vb \\
0 & 0 & \la_{321} v & \mltp & 0 & \la_{322} v \\
\la_{132} \vb & \la_{312} v & 0 & \la_{232} \vb & \la_{322} v & 0
\end{array} \right) $}.
\ee
Eq.(\ref{e2}) is based on two assumptions. The first, and crucial, condition
is that no $N-$scalar gets a vev much in excess of $10^9$ GeV; otherwise
nonrenormalizable operators of the form $\frac {1}{M_{Pl}} \langle N^* \rangle
\langle N^* \rangle \wt{N} \wt{N}$ could produce large entries on the diagonal
of the mass matrix, and the arguments presented below would be invalid. The
second, purely technical assumption is that the first two generations of
exotic ``leptons" do not mix with the gaugino--higgsino sector of the model.
This is technically natural \cite{11}, and requires that $\la_{33i} =
\la_{3i3} = \la_{i33} = 0$ for $i=1,2$. This second assumption has been made
purely for the sake of convencience; relaxing it will not change the result
significantly.

Note that some of the entries of the mass matrix (\ref{e2}) are equal to the
masses of the charged exotic ``leptons", which are given by $m_{L_i^\pm} =
\la_{ii3} x$. Since (in the absence of a signal for new gauge bosons) no upper
bound on the singlet vev $x$ can be given, these charged exotics can be
very heavy. Also, future \epem\ colliders will easily detect them unless their
masses exceed the beam energy. In contrast, all other entries of this mass
matrix are of order 100 GeV or less; otherwise at least one of the couplings
appearing in eq.(\ref{e1}) would have a Landau pole at a rather low energy
\cite{8}, and the theory would no longer be weakly interacting. If lower
bounds on the $m_{L_i^\pm}$ continue to increase, we will soon be in a
situation where the mass matrix (\ref{e2}) can be approximately diagonalized
analytically. There are then four heavy eigenstates, with masses $\pm
m_{L_1^\pm}$ and $\pm m_{L_2^\pm}$, respectively. Since $\det {\cal M}_{L^0}
\propto (x v \bar v)^2$, the remaining two eigenvalues must be quite small
\cite{10}:
\be \label{e3}
m_{L^0_{1,2}} \propto \la^2 v \bar v / m_{L_i^\pm}.
\ee
Note that the upper bound on these eigenvalues {\em de}creases as the mass of
the charged exotic ``leptons" is {\em in}creased \cite{8}.

This is illustrated in Fig.~1, taken from ref.\cite{7}, which shows the upper
bound on the mass of the lightest neutral exotic ``lepton" as a function of
the lower bound on the mass of the charged exotic ``leptons". Note that we
also have required $m_{L^0_1} + m_{L^0_3} > 2 m_{L^\pm_{\rm min}}$ here, so
that associate $L^0_1 L^0_3$ production is also kinematically suppressed. This
reproduces the upper bound on $m_{L_1^0}$ that can be derived from the
nonobservation of signals for exotic ``lepton" production, including a proper
treatment of all relevant cross sections \cite{7}.

Increasing the lower bounds on the masses of the charged, or heavy neutral,
exotic ``leptons" also decreases the $SU(2)$ doublet components of the light
eigenstates $L_1^0$ and $L_2^0$. This decreases their production cross
sections at \epem\ colliders, and thus makes it easier for them to escape
detection. However, it can also lead to cosmological problems. The reason is
that the
exotic ``leptons" have odd $R-$parity if they have any superpotential
couplings to ordinary matter. If $R-$parity is conserved, $L_1^0$ must
therefore be absolutely stable if it is lighter than the lightest neutralino.
Notice that in this class of models, the SUSY breaking scale is generally of
order of the mass of the new $Z'$ boson, {\em not} of order $m_Z$. Fig.~1
shows that \mlone\ will have to be smaller than the mass of the lightest
neutralino if a 1 TeV \epem\ collider fails to discover exotic ``leptons", and
if LEP2 fails to discover charginos.\footnote{The failure to discover
charginos implies a lower bound on the mass of the lightest neutralino, since
the ``singlino" will decouple from the other neutralinos in the limit $x \gg
v, \vb$.} A stable $L_1^0$ might be problematic, since some fraction of the
``leptons" that were produced during the Big Bang are still around today. This
fraction is inversely proportional to the total annihilation cross--section
for $L_1^0$ pairs \cite{12}:
\be \label{e4}
\Omega(L_1^0) h^2 \propto \frac{1} {\sigma \left( L_1^0 L_1^0 \to
{\rm anything} \right)}
\sim \left[ \frac {m^2_{L_1^0}}{m^4_{H,A,Z}} ( {\rm coupling \ constants})
\right]^{-1},
\ee
where $\Omega (L_1^0)$ is the present mass density of $L_1^0$ particles in
units of the critical (closure) density, and $h$ is the re--scaled Hubble
constant. The requirement that the Universe be at least 10 billion years old
implies $\Omega h^2 \leq 1$. The second approximate equality in eq.(\ref{e4})
assumes that $L_1^0$ is considerably lighter than the particles through the
exchange of which $L_1^0$ pairs can annihilate; these are the $Z$ boson, the
pseudoscalar Higgs boson $A$, and the three neutral scalar Higgs bosons. The
coupling of $L_1^0$ to these bosons always involves at least one factor of a
(small) $SU(2)$ doublet component of $L_1^0$; the annihilation cross section
therefore has two such factors.

Increasing the lower bounds on the masses of the heavy exotic ``leptons" thus
increases the $L_1^0$ relic density in two ways: By reducing (the upper bound
on) \mlone, and by reducing the doublet components of $L_1^0$, and hence its
couplings to gauge and Higgs bosons. The requirement $\Omega(L_1^0) h^2 \leq
1$ tends to exclude scenarios with light, singlet--like $L_1^0$ states, which
are very difficult to probe directly at \epem\ colliders, as noted above; it
therefore nicely complements exotic ``leptons" searches at future colliders.

This is illustrated in Fig.~2, which shows the fraction of parameter space
surviving certain constraints, as a function of \mlone. The dotted histogram
has been obtained by requiring that a future 1.5 TeV collider does not find
{\em charged} exotic ``leptons", and that LEP1 did not find the light neutral
exotics; this latter constraint turns out to be trivially fulfilled here. The
dashed histogram shows that searches for {\em neutral} exotics at a 1.5 TeV
collider can rule out many scenarios where the charged exotics escape
detection; this shows that an \epem\ collider is much better suited for this
test than a $\gamma \gamma$ collider of similar energy. However, most
parameter combinations with a light $L_1^0$ still survive. Finally, the solid
histogram shows that only a very small region of parameter space remains
allowed once the relic density constraint $\Omega h^2 \leq 1$ has been
imposed.

It turns out \cite{7} that these few surviving scenarios always
predict a large invisible branching ratio for at least one light neutral Higgs
boson. I argued above that the relic density constraint requires $L_1^0$ to
have a significant $SU(2)$ doublet component, and hence generically sizable
couplings to gauge and Higgs bosons. Further, unsuccessful searches for
neutral exotics at a 1.5 TeV collider, together with the constraints from
unsuccessful searches for Higgs bosons at LEP1, imply that $2 \mlone$ is
indeed below the mass of some light Higgs boson, so that Higgs$\to L_1^0
L_1^0$ decays are allowed. Detailed calculations \cite{7} show that these
decays overwhelm the otherwise dominant decays into $b \bar b$ and $W W^*$
final states, leading to an invisible branching ratio between about 50\% and
95\% for at least one, and possibly two, light Higgs boson(s). Such Higgs
bosons would be readily detectable at an \epem\ collider with $\rs \geq 300$
GeV, where their invisible branching ratio can also be measured to $\sim 10\%$
precision. Hence the model could be completely ruled out if a 300 GeV collider
fails to discover a light Higgs boson with large invisible branching ratio,
and a 1.5 TeV collider fails to discover signals for the production of exotic
``leptons". The only assumptions we have made are that $R-$parity is
conserved, and that no $N$ scalar gets a vev larger than $10^9$ GeV or so.

There is a possible loophole, however. This analysis relied heavily on the
form of the mass matrix for the neutral exotics, eq.(\ref{e2}); in particular,
the zeroes in the $(\wt{N_i},\wt{N_j})$ entries are crucial for the derivation
of eq.(\ref{e3}). These zeroes are a consequence of gauge invariance under the
new $U(1)$ and/or $SU(2)$ group factors characteristic for string--inspired
$E(6)$ models \cite{1}. However, we know that this additional gauge symmetry
must be broken quite badly, since the corresponding new gauge bosons have to
be heavy. It is therefore possible \cite{13} that one--loop corrections
produce new sizable non--zero entries in the mass matrix (\ref{e2}). The most
dangerous contributions come from diagrams where an $\wt{N_i}$ fermion splits
into an $\wt{H}$ fermion and an $\bar H$ boson (or vice versa), and then
re--combines (with a mass insertion on the internal fermion line) into an
$\wt{N_j}$ fermion. This would give a contribution of order
\be \label{e5}
m^{\rm 1-loop}_{\wt{N_i}\wt{N_j}} \sim \frac {\la^4} {8 \pi^2} A,
\ee
where \la\ stands for any coupling in eq.(\ref{e1}), and $A$ is a trilinear
soft SUSY breaking parameter, which appears because a nonzero contribution
necessitates mixing between $H$ and $\bar{H}$ scalars. Since the couplings in
eq.(\ref{e1}) all have to be $\leq 1$, the contribution (\ref{e5}) can only be
significant if $A$ is considerably larger than 1 TeV, which raises finetuning
problems. Note also that such a contribution, if sizable, will tend to reduce
the $SU(2)$ doublet components of the light neutral exotics, and hence their
couplings to gauge and Higgs bosons; this might exacerbate the cosmological
problems. A more complete calculation is necessary before we can decide
whether this ugly little loophole indeed exists.

\subsection*{Acknowledgements}
This talk is based on work done in collaboration with Atsushi Yamada. I thank
the governor of Iwate prefecture, the mayor of Ashiro, and the crows for their
warm hospitality and cold sake. This work was supported in part by the U.S.
Department of Energy under grant No. DE-FG02-95ER40896, by the Wisconsin
Research Committee with funds granted by the Wisconsin Alumni Research
Foundation, as well as by a grant from the Deutsche Forschungsgemeinschaft
under the Heisenberg program.

\end{document}